\documentclass[12pt]{iopart}
\usepackage{iopams} 
\usepackage{graphicx}
\usepackage{subfigure}
\usepackage{xcolor}
\usepackage{soul}
 \expandafter\let\csname equation*\endcsname\relax 
 \expandafter\let\csname endequation*\endcsname\relax 
\usepackage{amsmath}
\usepackage{wasysym}
\usepackage{amssymb}
\usepackage{color}
\usepackage{soul}
\usepackage{ulem}

\begin{document}

\title[Preferred degree networks]{Modeling interacting dynamic networks: II. Systematic study of the
statistical properties of cross-links between two networks with preferred
degrees }
\author{Wenjia Liu$^{1}$, B. Schmittmann$^{1,2}$, and R. K. P. Zia$^{1,2}$%
}

\address{$^1$Department of Physics and Astronomy, Iowa State University, Ames, IA 50011, USA}
\address{$^2$Department of Physics, Virginia Tech, Blacksburg, VA 24061}

\ead{wjliu@iastate.edu, schmittb@iastate.edu and rkpzia@vt.edu}

\begin{abstract}
In a recent work \cite{LiuJoladSchZia13}, we introduced dynamic networks with preferred degrees and presented simulation and analytic studies of a single, homogeneous system as well as two interacting networks. Here, we extend these studies to a wider range of parameter space, in a more systematic fashion. Though the interaction we introduced seems simple and intuitive, it produced dramatically different behavior in the single- and two-network systems. Specifically, partitioning the single network into two identical sectors, we find the cross-link distribution to be a sharply peaked Gaussian. In stark contrast, we find a very broad and flat plateau in the case of two interacting identical networks. A sound understanding of this phenomenon remains elusive. Exploring more asymmetric interacting networks, we discover a kind of `universal behavior' for systems in which the `introverts' (nodes with smaller preferred degree) are far outnumbered. Remarkably, an approximation scheme for their degree distribution can be formulated, leading to very successful predictions. 
\end{abstract}

%\pacs{ 05.10.Ln, 05.70.Ln, 05.90.+m,}
\maketitle
\section{INTRODUCTION}

Complex networks can be found everywhere in our world,
ranging from neuronal architectures to galactic filaments and from facebook
to transportation systems \cite
{Strogatz01,AlbertBarabasi02,DorogovtsevMendes02,Newman03,EstradaFoxHighamOppo10}%
. To understand the behavior of these systems requires both powerful
analytic tools and access to large data sets. Thanks to the rapid pace of
development in information technology, considerable amount of data can be
collected and systematic descriptions of various features of these systems
have been initiated. At the same time, an interdisciplinary academic field
-- network science -- has become quite mature, establishing a powerful
framework for characterizing and analyzing complex networks, as well as
applying network representations successfully to study physical, biological,
and infrastructure systems. In this context, early studies focus mainly on \textit{static} networks, and are quite adequate for describing networks
such as power grids and highways, whose topology can be regarded as constant
(within the time scales of interest). However, in many other situations,
such as social networks, a \textit{dynamic} description would be more
appropriate. Yet, there are far fewer studies for such networks. Recent and
notable examples include the time evolution of network topology 
\cite{WattStrogatz98,AlbertJeongBarabasi99,BarabasiAlbert99}, 
dynamical
processes on networks 
\cite{BarratBarthelemyVespignani08,DorogovtsevGAMJ08}, 
as well as the combination
of both, namely, adaptive co-evolutionary networks 
\cite{GrossDCBB06,GrossBlasius08}. 
Within the physics community, most of
these studies have focused on single isolated networks, putting aside the
fact that real world systems are highly interconnected and therefore should
be modelled as interacting networks. For instance, smartphones can help drivers avoid heavy traffic. 
This situation cannot be fully described in terms of a single network, whether we focus on cellular communication or the transportation (road) network. Therefore, models with interdependent
networks are needed, culminating perhaps, in a `theory of networks of
networks.' In recent years, the significance of interdependent networks has
begun to attract attention, and some aspects of such networks have been
probed. Those studies include the investigation of critical infrastructure
interdependencies \cite%
{RinaldiPeerenboomKelly01,PanzieriSetola08,Vespignani10,BuldyrevParshaniPaulStanleyHavlin10,BuldyrevShereCwilich11}%
, and approaches such as the multilayer method to couple traffic flows
to physical infrastructures \cite{KurantThiran06}.

While our ultimate goal is to understand the interdependence of dynamic
networks, we begin with simple model systems, in order to gain some insight
into the effects of interactions. In particular, we focus on networks with
`preferred degrees,' which allow us to implement both dynamics and
interactions easily. In the first paper of this series 
\cite{LiuJoladSchZia13}, 
we introduced such a dynamic network, in which each node
is pre-assigned a preferred degree ($\kappa $) and, when chosen to act,
adds/cuts links to reach and maintain $\kappa $. Since the dynamics does not
obey detailed balance in general, these systems settle eventually into 
\textit{non-equilibrium steady states} (in contrast to systems in thermal
equilibrium controlled by Boltzmann weights). In our studies here, we devote
attention solely to such states. We first consider a homogeneous
population and discover some unexpected properties of this network. Then, we
introduce a coupling between two such networks and
investigate the effects of their interaction. In \cite{LiuJoladSchZia13}, we focused predominantly on various degree distributions and found that they differ significantly from the
Poisson in a standard Erd\H{o}s-R\'{e}nyi random network \cite{E-R}. Within
the limited range of parameter space studied there, the degree distributions ($\rho(k)$'s) can be
reasonably explained by a mean-field approach. For a few special choices of parameters, we already noted that the interaction between just two populations can induce highly
non-trivial behavior of $X$, the total number of cross-links between them.
In this paper, we probe the parameter space more systematically, focusing specifically on the
properties of the mean, $\left\langle X\right\rangle $, and standard
deviation, $\sigma _{X}$. In some special cases, we delve into more detail,
such as various degree distributions in the steady state 
($\rho ^{ss}\left(k\right) $), as well as the distribution for 
the cross-links: $P^{ss}(X)$.
For a homogeneous
population, by defining $X$ to be the total number of cross-links between
two identical partitions, we find the stationary distribution $P^{ss}(X)$ to
be a narrowly peaked distribution, well described by a Gaussian. By
contrast, for a very similar two-network model, which we refer to as the
`symmetric system,' $P^{ss}(X)$ displays a broad and flat plateau! Moreover,
the power spectrum of $X\left( t\right) $ shows that the dynamics of $X$ is
consistent with an \textit{unbiased} random walk (within some bounds). The
dramatic difference between such similar models illustrates that the
non-trivial behavior of $X$ is indeed a consequence of the interaction between networks. 

The remainder of this paper is organized as follows. In
Section 2, we present the specifications of our models and introduce several
quantities that serve to describe the topology of our networks. In Section 3,
we show the Monte Carlo simulation results along with some analytical
understandings. In the last Section, we provide a summary and outlook for
this paper.

\section{SPECIFICATIONS OF THE MODELS}

\subsection{A single network with a preferred degree (single-network model)}

Recently, we introduced a class of dynamic networks evolving according to
one or more preferred degrees \cite{PlatiniZia10,ZiaLiuJoladSchmitt11,
JoladLiuSchmittmannZia12,LiuSchmittmannZia12,LiuJoladSchZia13}. The
motivation of such a model lies with the belief that, in typical social
settings, an individual would \textit{prefer} to have a certain number of
contacts. For example, an introvert may prefer only a handful of friends,
while an extrovert would be glad to have hundreds or thousands of contacts. Of course, real social interactions are far more complex, and so we use the notion of `extroverts' and `introverts' only as an illustration.
For the readers' convenience, we briefly summarize the main features of our
model here.

In the simplest case, we model a homogeneous population of $N$ nodes
(individuals), all assigned the same `preferred degree,' $\kappa $. When
chosen for updates, a node will attempt to add or cut one of its links based
on $\kappa $. In each attempt, a node is randomly chosen, its degree ($k$)
is noted, and depending on whether $k$ is smaller/larger than $\kappa $, $k$
will be increased/decreased by one. (To avoid ambiguity, $\kappa $ is always
chosen to be slightly larger than the integer quoted, e.g., $\kappa =25$
really means $25.5$. Here, this step is deterministic, while stochastic
rules can be implemented \cite{ZiaLiuJoladSchmitt11,LiuJoladSchZia13}.) In
this sense, a node `prefers' to have degree $\kappa $. For simplicity, the
action of adding/cutting is performed on a randomly chosen partner which has
no/a link with the node. The partner node is passive and has no influence on
this action. Self-loops and multiple connections are not allowed. In our
simulation, one Monte Carlo step (MCS) consists of $N$ such attempts, so
that, on the average, each node has one chance to take action. Clearly, the
network is dynamic, while the node attributes remain static. At large times,
the system will reach a steady state, with statistically stationary network
topology. Not surprisingly, the average degree is $\kappa $. However,
despite the appearance of randomness, the degree distribution is neither
Poisson nor Gaussian, but Laplacian \cite{LiuJoladSchZia13}.

\subsection{Modelling the interaction between two networks (two-network model)%
}

Our goal is to study interactions between two such networks, with different $%
\kappa $'s and $N$'s in general. One quantity of interest is $X$, the total
number of cross-links between the two networks. $X$ is clearly a
quantitative measure of the interaction between them. Next, let us introduce
the `interactions.' Of course, there are infinitely many ways to do so and
we can explore only a few, motivated by what seems the most likely behavior
between individuals in two populations. In this paper, we begin with
arguably the simplest: $\chi $, the probability a node acts (add or cut) on
a cross-link. A few other forms of natural `interactions' will be considered
in the last paper of this series.

Consider two preferred degree networks (labelled by $\alpha =1,2$), with $%
N_{\alpha }$ nodes, preferred degrees $\kappa _{\alpha }$, and cross-link
action $\chi _{\alpha }$. In each attempt, one node is chosen at random from
all the $N$ ($=N_{1}+N_{2}$) nodes. If the degree of this node is lower/higher than $\kappa _{\alpha }$, it will attempt to add/cut a
link. With probability $\chi _{\alpha }$, this action will be taken with a
partner node from the \textit{other} network. Thus, an intra-community link
will be updated with probability $1-\chi _{\alpha }$. In all cases, the
partner node will be randomly picked from the chosen group. If a suitable
partner does not exist (e.g., when the action is to cut and there are no
links to nodes in the chosen community), then no action is taken. As usual,
one MCS involves $N$ such attempts. In all our simulations, the initial
network is entirely devoid of links, i.e., a null graph.

With this set-up, the parameters $\chi _{\alpha }$ clearly control the
behavior of $X$. In the extreme case of $\chi _{\alpha }=0$, $X\equiv 0$ as
the two networks decouple completely. At the other extreme, $\chi _{\alpha
}=1$, the system consists of only bipartite graphs, though not a complete
one in general ($X<N_{1}N_{2}$). With only cross-links, such a system may be
regarded as `fully interacting.' In this sense, $\chi $ plays the role of an
interaction strength.

\subsection{Quantities of interest}

One of the standard characterizations of the topology of a network is the
degree distribution, $\rho (k)$. Denoting by $n_{k}$ the number of nodes
with $k$ links in each measurement, $\rho (k)$ is given by 
\begin{equation}
\rho (k)=\frac{\langle n_{k}\rangle }{N}.
\end{equation}
For a homogeneous network with a single preferred degree, this $\rho $ is,
as expected, sharply peaked around $\kappa $. In a system with two
sub-networks with different preferred degrees, it is expected to be bimodal,
especially if the $\kappa $'s are far apart. Thus, it is sensible to
consider separate distributions, $\rho _{\alpha }\left( k\right) $%
, associated with nodes in community $\alpha $ which have degree $k$%
. Beyond these, we may extend our considerations to the next level of
detail, $\rho _{\alpha \beta }\left( k_{\alpha \beta }\right) $, associated
with $k_{\alpha \beta }$, denoting the number of links with which a node in community $\alpha $
is connected to nodes in community $\beta $. Note that $\left\langle
k_{12}\right\rangle \neq \left\langle k_{21}\right\rangle $, since the
average number of cross-links $\left\langle X\right\rangle $ is equal to
both $N_{1}\left\langle k_{12}\right\rangle $ and $N_{2}\left\langle
k_{21}\right\rangle $. Though somewhat cumbersome (compared to $\rho _{\alpha \beta }\left( k\right) $), we
will use the notation above, to leave no doubt about which quantity is
being considered. We will also refer to $\rho _{\alpha }$ as a `global
degree distribution,' reserving the terms internal/external degree
distribution for $\rho _{\alpha \alpha}$ / $\rho _{\alpha \beta }$ . Our study
will be mainly for the steady state, for which we add the superscript $ss$,
e.g., $\rho _{12}^{ss}$.

Of course, we can proceed further and consider joint distributions, such as $%
P_{1}\left( k_{11},k_{12}\right) $, the probability that a node in community 
$1$ will be found with degrees $k_{11}$ \textit{and} $k_{12}$, etc. For this
paper, however, we will limit ourselves to the less detailed distributions.
Clearly, 
\begin{equation}
\rho _{\alpha }\left( k\right) \equiv \sum_{k_{\alpha \alpha
},k_{\alpha \beta }}\delta \left( k_{\alpha \alpha }+k_{\alpha \beta
}-k\right) P_{\alpha }\left( k_{\alpha \alpha },k_{\alpha \beta
}\right)
\end{equation}
while $\rho _{\alpha \alpha}$ and $\rho _{\alpha \beta }$\ are simple projections
of $P_{\alpha }$, e.g., $\rho _{\alpha \alpha}\left( k_{\alpha \alpha }\right)
=\sum_{k_{\alpha \beta }}P_{\alpha }\left( k_{\alpha \alpha },k_{\alpha
\beta }\right) $. These remarks show that $\rho _{\alpha }$, e.g., cannot be
obtained from $\rho _{\alpha \alpha}$ and $\rho _{\alpha \beta }$ in general.

As we are interested in the behavior of the cross-network interactions, most
of our attention will be on cross-links. While much information is stored in 
$\rho _{\alpha \beta }$, it is more efficient to study the `macroscopic'
quantity, $X$, specifically its mean, $\left\langle X\right\rangle $, and
standard deviation, $\sigma _{X}$. For a special case (see Eqn.~(\ref{Symmetric})
below), its time dependence in the steady state will be analyzed in more
detail. Thus, $X(t)$ will be used to compile a histogram which represents $%
P^{ss}(X)$, while its power spectrum will be exploited to reveal the nature
of the dynamics leading to $P^{ss}$. In particular, a surprising discovery
is that, under certain conditions, $X$ performs an unbiased random walk over
an extremely large fraction of its available range, so that $P^{ss}$
displays a broad plateau instead of a sharp peak.

To determine which aspects of the interaction are crucial for the emergence
of such remarkable phenomena, we compare two very similar systems. One is a
homogeneous network of $2L$ nodes with $\kappa $. Arbitrarily labelling half
of them as `red' and the rest `blue,' we mimic having two
communities and define $X$ as the total number of `red-blue' links. The
other system is the `symmetric' two-network model: 
\begin{equation}
N_{1}=N_{2}=L;~~\kappa _{1}=\kappa _{2}=\kappa ;~~\chi _{1}=\chi _{2}=0.5
\label{Symmetric}
\end{equation}
The apparent symmetry between the two communities may lead us to expect that
the behavior of $X$ in this system should be similar to that in the
homogeneous population. As will be shown in the next Section, this naive
expectation is far from fulfilled and the simple interaction associated with 
$\chi $ has a profound effect on the macroscopic $X$.

Finally, in a social network with introverts and extroverts, `frustration'
is unavoidable. Instead of a qualitative notion, `frustration' can be
quantified in our model. For example, consider $\kappa _{1}\ll \kappa _{2}$.
Since a node plays a passive role for much of the time (as other individuals
add/cut links to it), we should expect an introvert/extrovert (characterized
by $\kappa _{1,2}$) to find itself mostly with more/less contacts than it
prefers. In the steady state, we may define `frustration' for \textit{%
individual} $i$ by 
\begin{equation}
\phi _{i}\equiv \sum_{k>\kappa }\rho ^{\left( i\right) }\left( k\right)
-\sum_{k<\kappa }\rho ^{\left( i\right) }\left( k\right)
\end{equation}
where $\rho ^{\left( i\right) }\left( k\right) $ is the degree distribution of $%
i $ alone (and not of the population as a whole). With $\phi \in \left[ -1,1\right] 
$, its magnitude is a measure of how frustrated $i$ is, while the sign
provides its propensity to add/cut in its attempt to seek relief. Although
we will not study this quantity in detail, we will use the concept in later
discussions. Specifically, in the third paper of this series \cite%
{LiuBasslerSchmittmannZia14}, we will investigate a `maximally frustrated'
population of extreme introverts ($\kappa _{1}=0_{-}$) and extroverts ($%
\kappa _{2}=\infty $). Preliminary results of this system, reported
elsewhere \cite{LiuSchmittmannZia12}, showed the existence of a sharp
transition in $X$ when $N_{1}-N_{2}$ changes sign.

\section{SIMULATION RESULTS AND THEORETICAL CONSIDERATIONS}

In general, there are three pairs of relevant control parameters when
coupling two preferred degree networks with $\chi $, namely, $N_{1,2}$, $%
\kappa _{1,2}$, and $\chi _{1,2}$. Exploring such a 6-dimensional space is
beyond the scope of our study. We content ourselves with a limited region in
certain subspaces, as most of our simulations focus on $N_{1}+N_{2}=200$, $%
\kappa _{1}+\kappa _{2}=50$, and $\chi _{1}+\chi _{2}=1$. In particular, we
begin with a symmetric system, Eqn.~(\ref{Symmetric}), and contrast its
behavior to that in a homogeneous network with $N=200$ and $\kappa =25$. For
more general cases, we reported results for systems with only one of the
three pairs being different (i.e., along certain 1-d subspaces) \cite
{LiuJoladSchZia13}. Here, we extend our studies to certain 2-d subspaces,
with two pairs of control parameters being distinct. Though an overall
understanding of the behavior of these networks remains beyond our grasp, we
are able to gain some insight, through a mean-field treatment, into systems
where one community is `fully frustrated.'

\subsection{Homogeneous \textit{vs.} symmetric heterogeneous populations}

A good baseline study of an interacting two-network system is the symmetric
case. Specifically, we performed simulations mostly with $N_{1}=N_{2}=100$, 
$\kappa _{1}=\kappa _{2}=25$, and $\chi _{1}=\chi _{2}=0.5$. Before we
proceed to the results, let us recapitulate, for comparison, what is known
about a very similar system consisting of just one homogeneous network 
($N=200$, $\kappa =25$). Though we find an expected result -- both systems
displaying the same degree distribution, $\rho ^{ss}\left( k\right) $, we
also discover a surprising one -- drastically different behavior in $X$.

\subsubsection{Steady state global degree distributions}

\begin{figure}[tbp]
\includegraphics[width=3.5in]{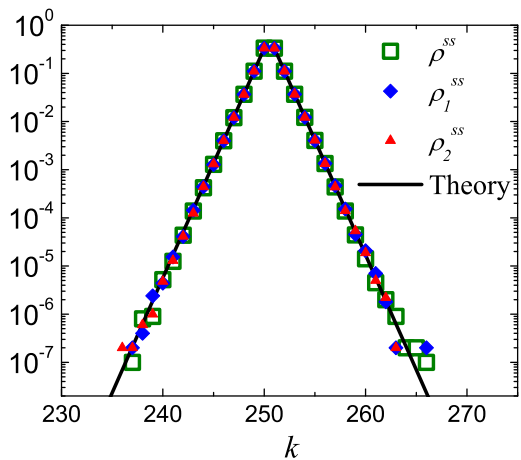} \centering
\caption{ Degree distribution, $\protect\rho ^{ss}$, of a single network 
with $N=1000$ and $\protect\kappa =250$ (green squares), along with the 
theoretical prediction (solid black line). For a system with two interacting
networks, the total degree distributions are $\protect\rho _1^{ss}$ 
(blue diamonds) and $\protect\rho _2^{ss}$ (red triangles).}
\label{DD}
\end{figure}

For the homogeneous population in the steady state, there is only one $\rho
^{ss}(k)$, since all links are connected to nodes in the same network. Of
all the distributions introduced above for two interacting networks, the
most appropriate ones for this comparative study is the global distribution, 
$\rho _1^{ss}$, which should be same as $\rho _2^{ss}$ for this symmetric
case. As we found, these quantities settle down quite rapidly, so that we
were able to exploit the relatively large systems ($L=1000$ and $\kappa =250$%
) used in previous studies \cite{LiuJoladSchZia13}. Starting with empty
networks, we discard the first $1K$/$2K$ MCS for the single-/two-network
model. Thereafter, we measure the quantities of interest every $100$ MCS and
compile the average over $10^4$ measurements. The resultant $\rho ^{ss}$'s
are shown in Fig.~\ref{DD}. It is clear that there is no discernible
difference between all three distributions, as might be expected. 
For the reader's convenience, we recapitulate known properties of $\rho
^{ss} $ and how its form (neither Poisson nor Gaussian) can be understood 
\cite{LiuJoladSchZia13}. We see that $\rho ^{ss}\propto e^{-\mu \left|
k-\kappa \right| }$ is a double exponential, i.e., a Laplacian. Our data
indicate $\mu =1.08\pm 0.01$, which can be explained using a crude
mean-field approximation to estimate the rates for our node to gain or lose
a link. Respectively, these are $\Theta (\kappa -k)+1/2$ and $\Theta
(k-\kappa )+1/2$, where $\Theta$ is the Heavyside function. The $\Theta $ terms correspond to how a chosen node will
act, while the $1/2$ accounts for how its partner nodes will act. In the
steady state, each node is `content' on the average, and so, the probability
to add/cut is just $1/2$. Balancing the gain/lose probability currents, we
arrive at 
\begin{equation}
\frac{\rho ^{ss}(k+1)}{\rho ^{ss}(k)}=\frac{\Theta (\kappa -k)+1/2}{\Theta
(k+1-\kappa )+1/2}
\end{equation}
which leads to $\rho ^{ss}(k)\propto 3^{-\left| k-\kappa \right| }$, in
excellent agreement with the data.

Needless to say, the same argument can be advanced for the \textit{symmetric}
two-network system, arriving at a similar result. However, as presented
below, drastic differences between the single-network and the interacting
networks emerge when we measure another quantity: $X$.

\subsubsection{Behavior of the total number of cross-links, $X$}

In a previous paper \cite{LiuJoladSchZia13}, we investigated this quantity
briefly. In particular, we found that $X(t)$ displays very large
fluctuations and, for systems with $L=1000$, it takes very long times ($\gg
3\times 10^6$ MCS) for $X$ to reach the limits of its range. To build reliable histograms, it would
take even longer to collect enough data. Thus, we consider smaller systems
for the remainder of this paper, namely, $L=100$ and $\kappa =25$. We also
discard the initial $10^{7}$ MCS, to let the system reach steady state,
before taking measurements every $100$ MCS for the next $2\times 10^{7}$
MCS. With the resultant time trace of $2\times 10^{5}$ points, we compile a
histogram, which leads to $P^{ss}(X)$. We also construct a power spectrum.
Specifically, we divide the entire time trace into 20 shorter ($%
10^{4}\equiv T$) ones and obtain Fourier transformations of each: $\tilde{X}%
(\omega )\equiv \sum_{t=1}^{T}{X(t)e^{i{\omega }t}}$ where $\omega =2{\pi }%
m/T~$($m{\in }[0,T-1]$). The power spectrum, $I(\omega )$, is defined as the
average $\langle |\tilde{X}(\omega )^{2}|\rangle $ over these 20 FT's.

\begin{figure}[tbp]
\centering
\includegraphics[width=3.5in]{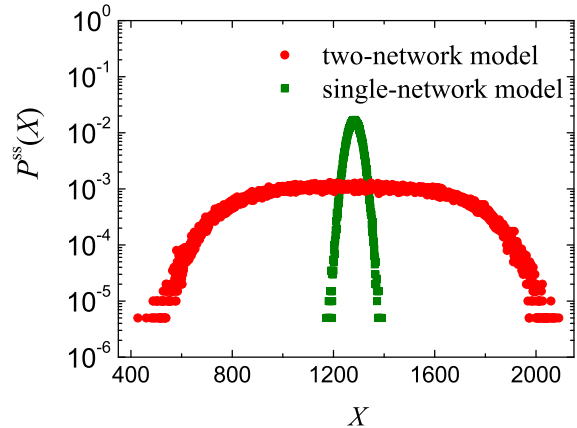}
\caption{Comparison of histograms for the number of cross-links $X$: from a single, 
homogeneous network with $N=200$, arbitrarily partitioned into two 
identical sections (green), and from a model with two interacting networks
with $N_{1}=N_{2}=100$ and $\protect\chi _{1}=\protect\chi _{2}=0.5$ (red). 
In all cases, the preferred degree, $\protect\kappa$, is $25$.}
\label{histogram}
\end{figure}

\begin{figure}[tbp]
\centering
\includegraphics[width=3.5in]{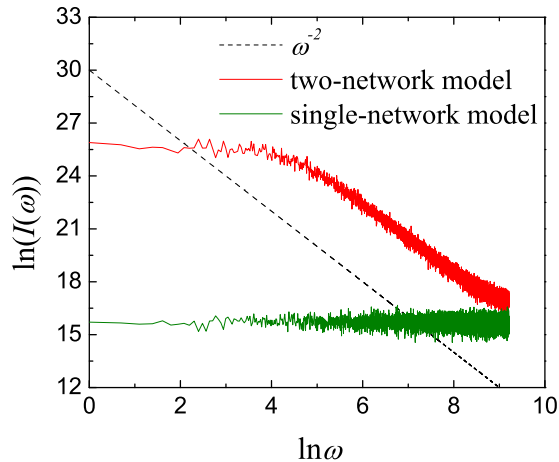}
\caption{Power spectra, $I(\protect\omega)$, associated with $X(t)$ for the
two systems described in the caption of Fig.~\ref{histogram}: 
green for the homogeneous network and red for the interacting two-network
model.}
\label{power}
\end{figure}

The results are plotted (in red) in Figs.~\ref{histogram} and \ref{power}.
The presence of a broad plateau in $P^{ss}(X)$ motivates us to explore the
dynamics of ${X(t)}$, to see if it simply performs an unbiased random walk
within the confines of two `soft walls.' This conjecture is confirmed by the
latter plot, in which we see that $I(\omega )$ indeed follows $1/\omega ^{2}$
quite well, crossing over (for small $\omega $) to a constant dictated by
the limits of $X$, namely, $0$ and $L^{2}$.

In stark contrast, the data displayed in green are indicative of very
different behavior. Let us emphasize what $X$ is -- in this homogeneous
population of $200$ nodes, all preferring degree $25$. First, we randomly
partition the system into 2 sets of 100 nodes, labelling one set `blue' and
the other `red.' At any $t$, the total number of `red-blue' links is defined
as $X$. Clearly, if we focus on any one node, there can be up to 199
contacts, though $\kappa $ will limit the average to about $25.5$. Thanks to
homogeneity and the randomness in the dynamics, we can expect half of these
to be cross-links. Thus, we are not surprised by the peak of $P^{ss}(X)$
being located at $\sim 100\times 25.5/2=1275$. We can go further, to
estimate the observed standard deviation ($\sigma _{X}\sim 25$) by the
following crude argument. Denoting by $x$ the number of cross-links of any
node, we already arrived at its mean, i.e., $\langle x\rangle =12.75$. If we
assume that the distribution of $x$ is a binomial distribution (with
probability 1/2 that the node acts on a cross-link), then this standard
deviation is, roughly, $\sigma _{x}=\sqrt{25.5/4}$. Invoking the central
limit theorem for 100 nodes, we easily find $\sigma _{X}=\sqrt{100}\sigma
_{x}\simeq 25$, in excellent agreement with observation. As for its
dynamics, we expect $X\left( t\right) $ to be governed by white noise, as
confirmed by the green line in Fig.~\ref{power}.

By contrast, $\sigma _{X}$ for the symmetric two-network system is over an
order of magnitude larger: $\sim 300$! (Of course, by symmetry, the
average $\left\langle X\right\rangle $ is expected to be the similar, i.e., $%
\sim 1250$.) The reason behind the failure of crude arguments for the
symmetric, interacting two-network system is quite subtle. Exploring this
non-trivial issue will be the topic of the third paper in the series \cite
{LiuBasslerSchmittmannZia14}. There, we will study the opposite (indeed,
extreme) limit of this system, thereby bringing the essence of this
remarkable behavior into sharp focus. For the remainder of this paper, we
will report on more systematic studies of $X$ in typical asymmetric
two-network models.

\subsection{Asymmetric heterogeneous populations}

In this subsection, we venture from the special symmetric system and explore
systematically a larger region of the 6-d parameter space: $\left( N_{\alpha
},\kappa _{\alpha },\chi _{\alpha }\right) $. As a result, we will be unable
to study details like the full $P^{ss}(X)$ or $I\left( \omega \right) $.
Instead, we will show results on just the mean, $\left\langle X\right\rangle 
$, and the standard deviation, $\sigma _{X}$.\footnote{
In our previous study \cite{LiuJoladSchZia13}, we considered only a few
specific points in this large parameter space. Thus, we were able to
investigate more detailed properties, such as the four $\rho _{\alpha \beta
} $'s.} To orient the reader in this space, we start from the specific case
above ($N_{\alpha }=100,\kappa _{\alpha }=25,\chi _{\alpha }=0.5$) and first
extend along the 1-d lines by varying just one of these pairs, but keeping
their values equal. While is it fair to refer to, say, a system with $\chi
_{1}=\chi _{2}=0.9$ as asymmetric (since the individuals do not choose
intra-community partners with the same probability as cross-links), the
other two `axes' represent genuinely symmetric systems. Those studies
(varying $N_{\alpha }$ or $\kappa _{\alpha }$ alone) should be regarded as
explorations of the effects of population size and degree preference. In all
these simulations, we again start with empty systems and carry out two
independent runs, each $10^{7}$ MCS long. $X$ is measured once every $100$
MCS, so that, for each case, there are $2\times 10^{5}$ data points from
which we compute the mean and standard deviation: $\langle X\rangle $, $%
\sigma _{X}$.

\subsubsection{Results from varying one pair of parameters.}

\begin{figure*}[tbp]
\centering
\mbox{
    \subfigure{\includegraphics[width=3in]{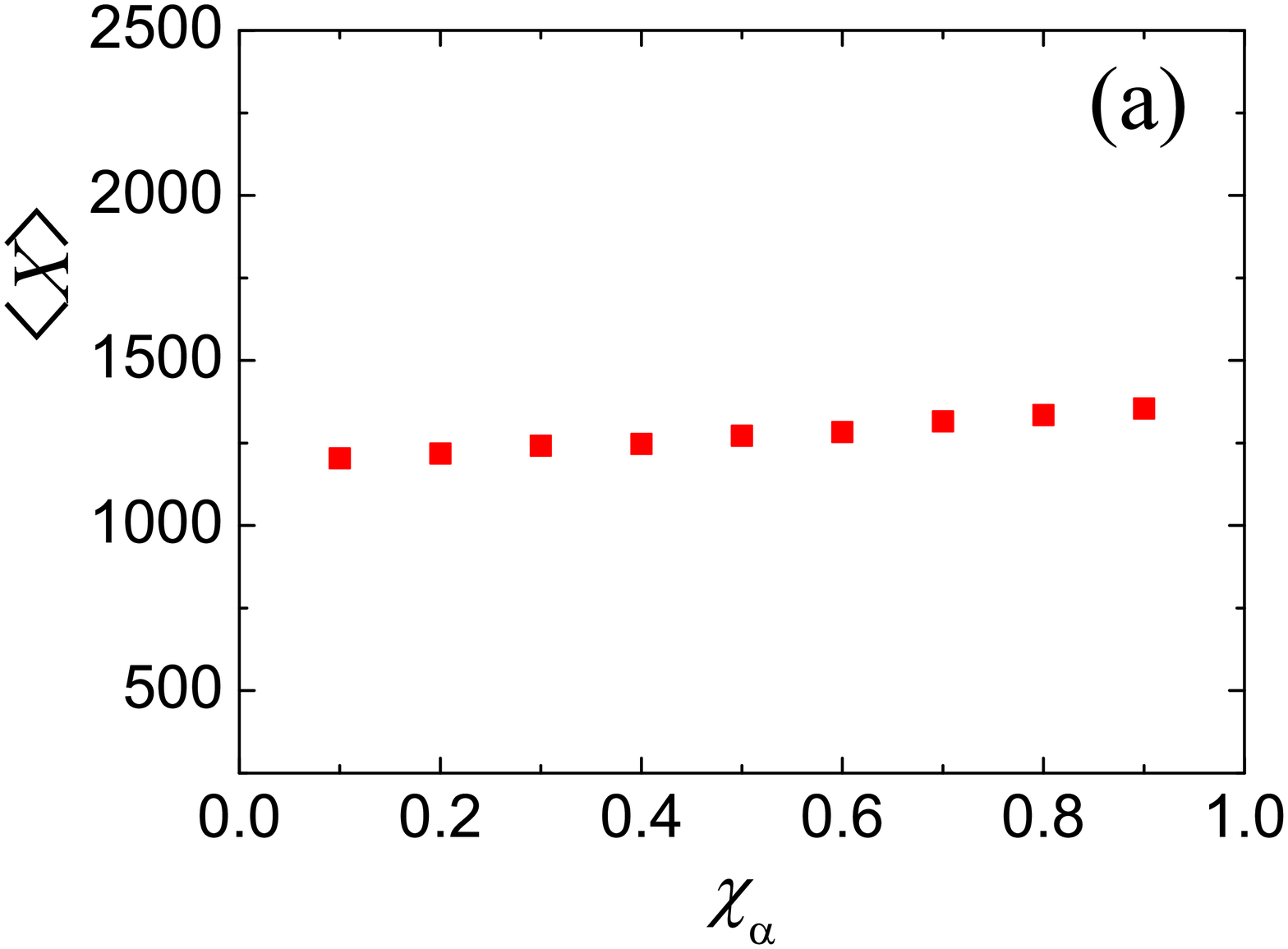}}\quad
    \subfigure{\includegraphics[width=3in]{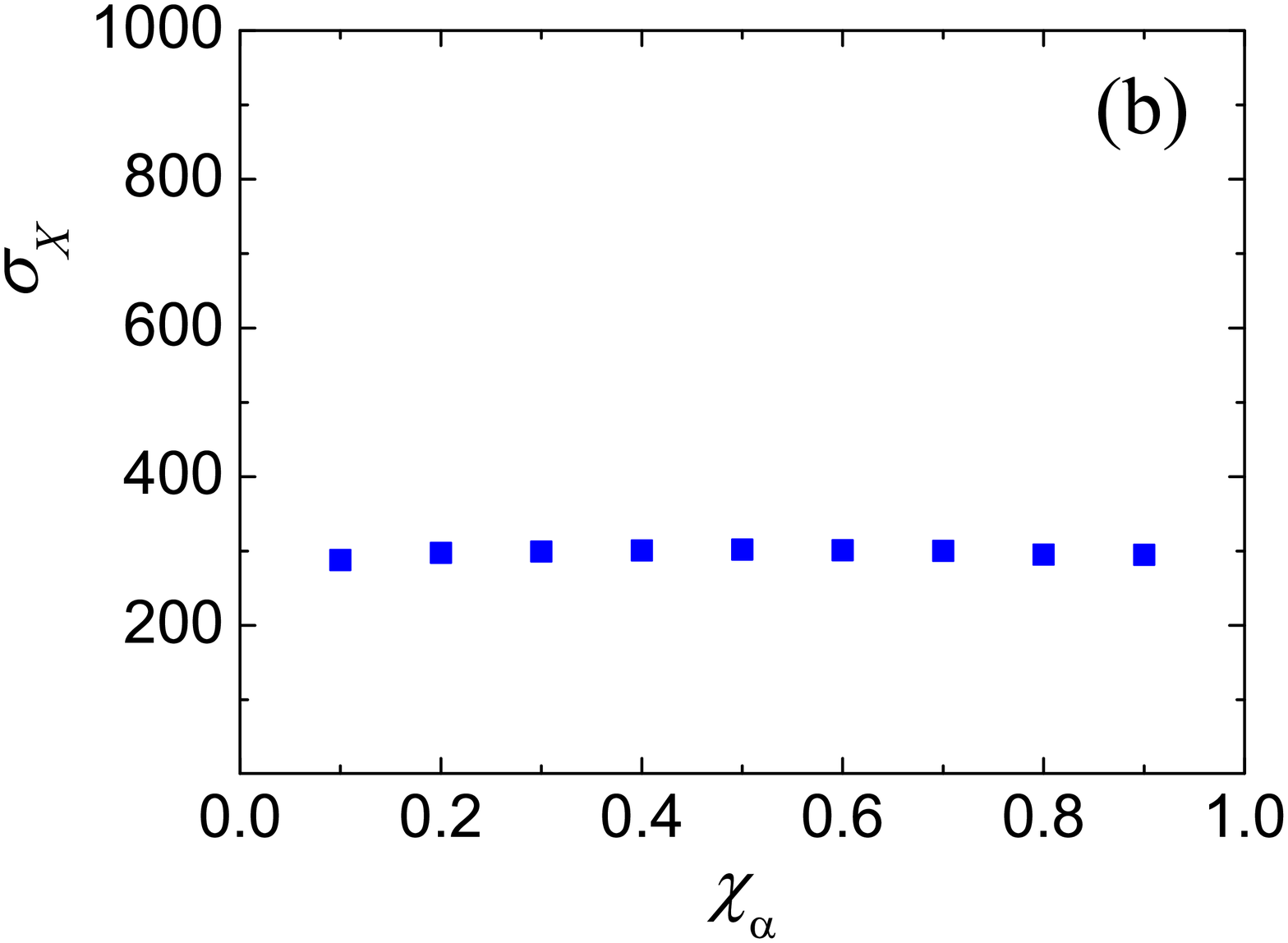}}
     }
\caption{The means (red squares) and standard deviations (blue squares) associated with cross-links 
 in a two-network model with $N_{\protect\alpha }=100$ and 
$\kappa _{\protect\alpha }=25$, as a function of 
$\protect\chi _{\protect\alpha }=\protect\chi _{1}=\protect\chi _{2}$.}
\label{diffS}
\end{figure*}

We begin by varying $\chi _{1}=\chi _{2}=\chi $, with $N_{\alpha }$
and $\kappa _{\alpha }$ kept at $100$ and $25$, respectively. Recall that $%
\chi $ controls effectively the interaction between two networks, since
individuals with larger $\chi $ are more likely to take action on
cross-links. Thus, $\chi =0$ represent two independent networks, while only
cross-links are present in $\chi =1$ networks. For general $\chi $, however,
it is not directly related to the value of $X$ (but only to the likelihood
for $X$ to change). Therefore, our expectation is that changing $\chi $
would not affect $\langle X\rangle $ or $\sigma_X $. Fig.~\ref{diffS} shows
the simulation results for $\chi =0.1,...,0.9$. Our expectation is largely
borne out, especially for $\sigma _{X}$. There appears to be a slight rising
trend in $\langle X\rangle $, $\sim 15\%$ over this entire range of $%
\chi $. It is difficult to explain these variations in detail, though the
typical values deviate little from that in the symmetric case ($\chi
=0.5;~~\langle X\rangle \sim 1272$), as predicted above. We will next see
that more interesting behavior appears when we vary the other two control
parameters. 

\begin{figure*}[tbp]
\centering
\mbox{
    \subfigure{\includegraphics[width=3in]{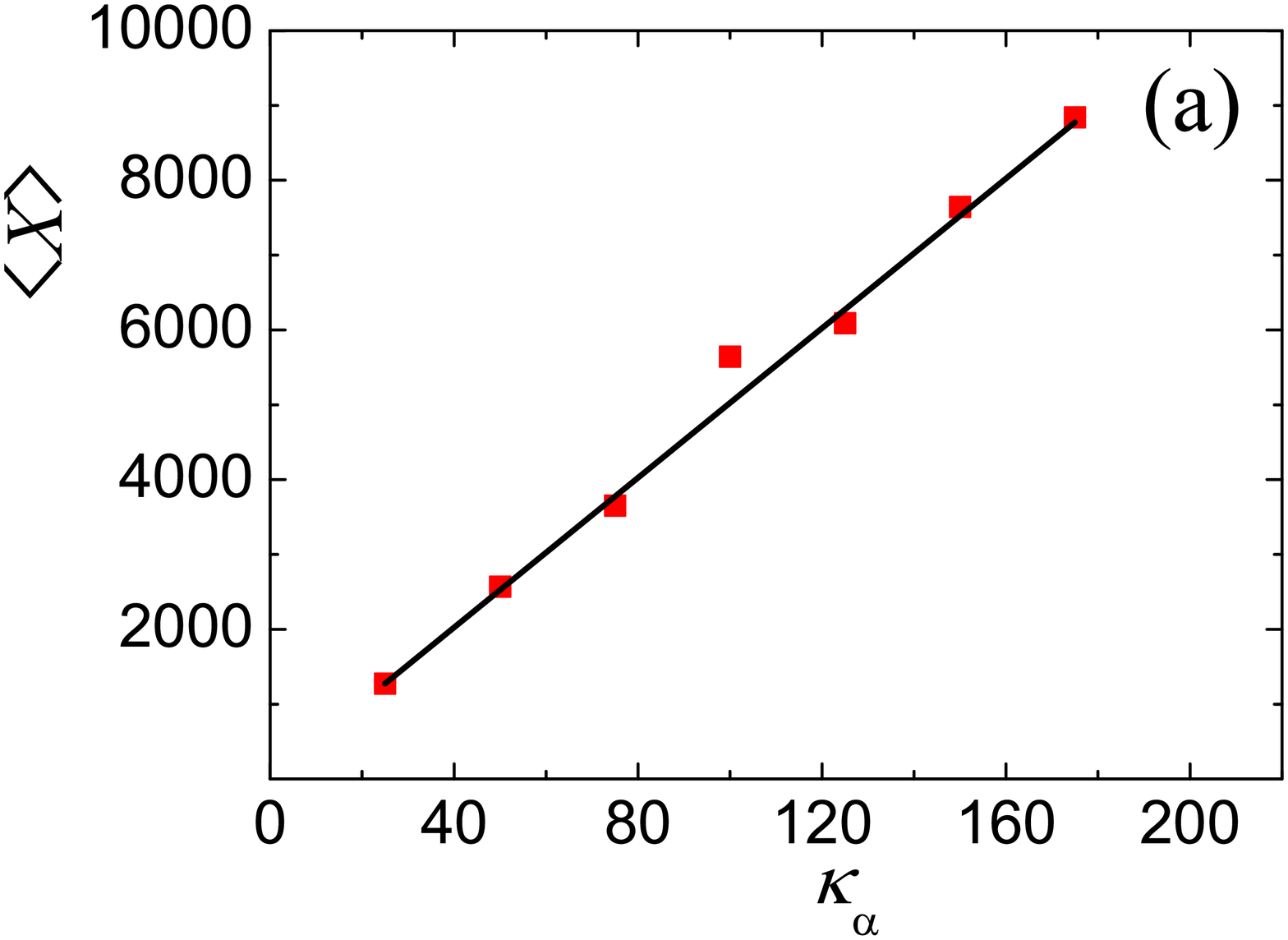}}\quad
    \subfigure{\includegraphics[width=3in]{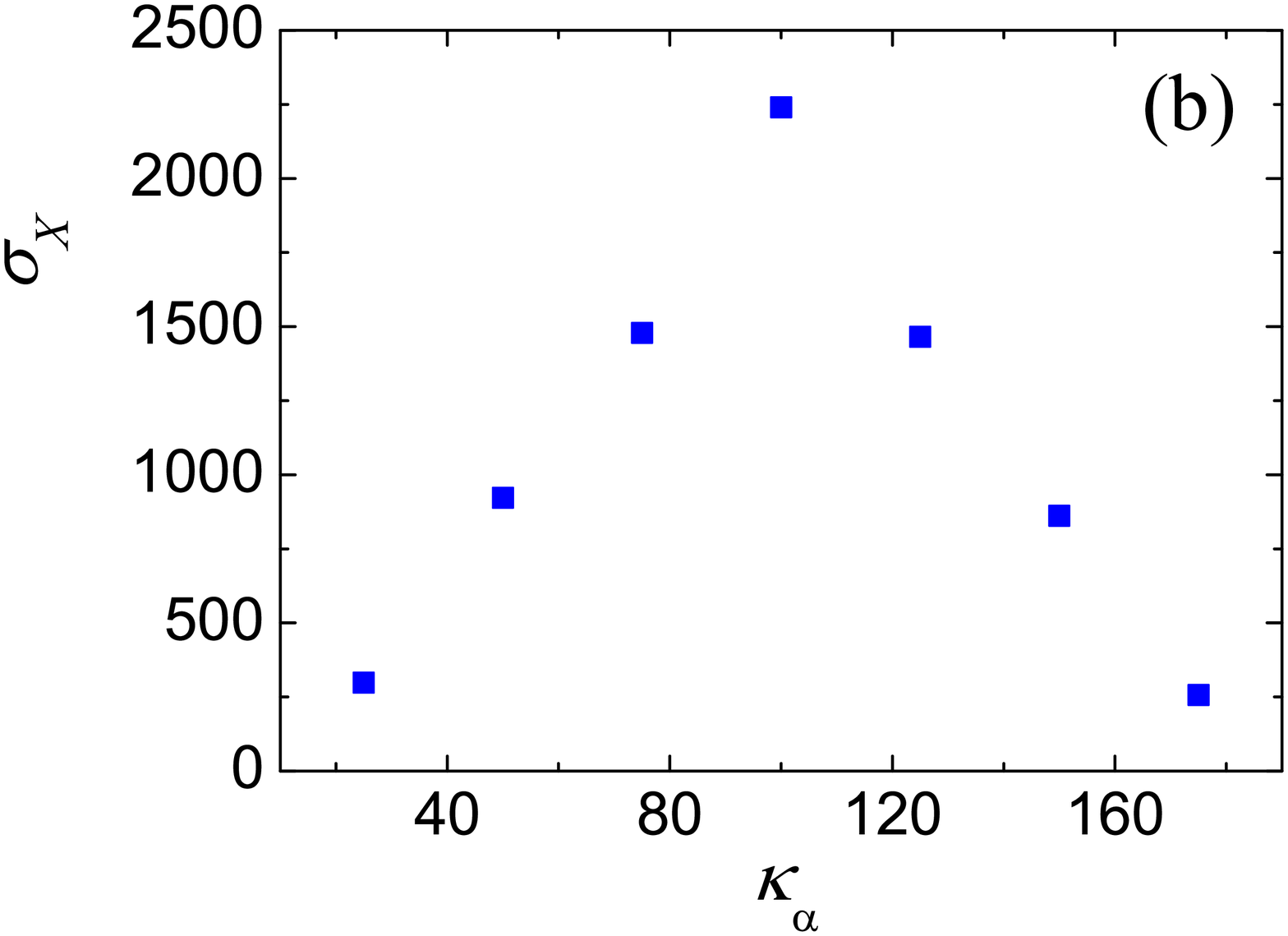}}
     }
\caption{The means (red squares) and standard deviations (blue squares) associated with cross-links 
 in a two-network model with $N_{\protect\alpha }=100$ and 
$\protect\chi _{\protect\alpha }=0.5$, as a function of 
$\protect\kappa _{\protect\alpha }=\protect\kappa _{1}=\protect\kappa _{2}$. In (a), $N_\alpha\kappa_\alpha/2$ is also plotted (solid line) for comparison.}
\label{diffk}
\end{figure*}

The next pair we vary is $\kappa _{1}=\kappa _{2}=\kappa $, with fixed $%
N_{\alpha }=L=100$ and $\chi _{\alpha }=\chi =0.5$. Fig.~\ref{diffk} shows
the simulation results for $\kappa =25,50,75,...,175$. Needless to say, if $%
\kappa $ exceeds the total population size, ($200$ in simulations here),
every link will be established quickly, and $X$ will be a constant $L^{2}=10^{4}$
and $\sigma _{X}\equiv 0$. Thus, our simulations sample essentially the
entire range of meaningful $\kappa $'s here. Now, $\kappa $ controls the
preferred degree of a node and so, it is not surprising that $\langle
X\rangle \propto \kappa $. In particular, since $\rho _{\alpha }(k)$ is
sharply peaked at $\kappa _{\alpha }$, a simple minded estimate is that each
node has $\kappa _{\alpha }\chi _{\alpha }$ cross-links. Thus, we arrive at $%
\langle X\rangle \simeq \kappa _{\alpha }N_{\alpha }/2$ ($=50\kappa $ here),
plotted as a solid line in Fig.~\ref{diffk} and in surprisingly good
agreement with the data. We are unable to find a similar estimate for the
behavior of $\sigma _{X}$. The most striking features there are: (i) $\sigma
_{X}$ varies linearly with $\kappa $ and (ii) $\sigma _{X}$ is symmetric
around its peaks at $100$. That $\sigma _{X}$ vanishes at both end points is
clear, but it is unclear how other features arise. A phenomenological
formula which accounts for these features is $\sigma _{X}\propto $ $\min
\left( \langle X\rangle ,L^{2}-\langle X\rangle \right) $, but how it
emerges from the underlying dynamics is unknown. In stark contrast, since
this system is also `symmetric,' an argument similar to the above (for its
counterpart in a homogeneous population) would provide $\sqrt{N\kappa /4}=%
\sqrt{25\kappa }$, which clearly fails to match the data. Understanding the
properties of $\sigma _{X}$ here remains a challenge.

\begin{figure*}[tbp]
\centering
\mbox{
    \subfigure{\includegraphics[width=3in]{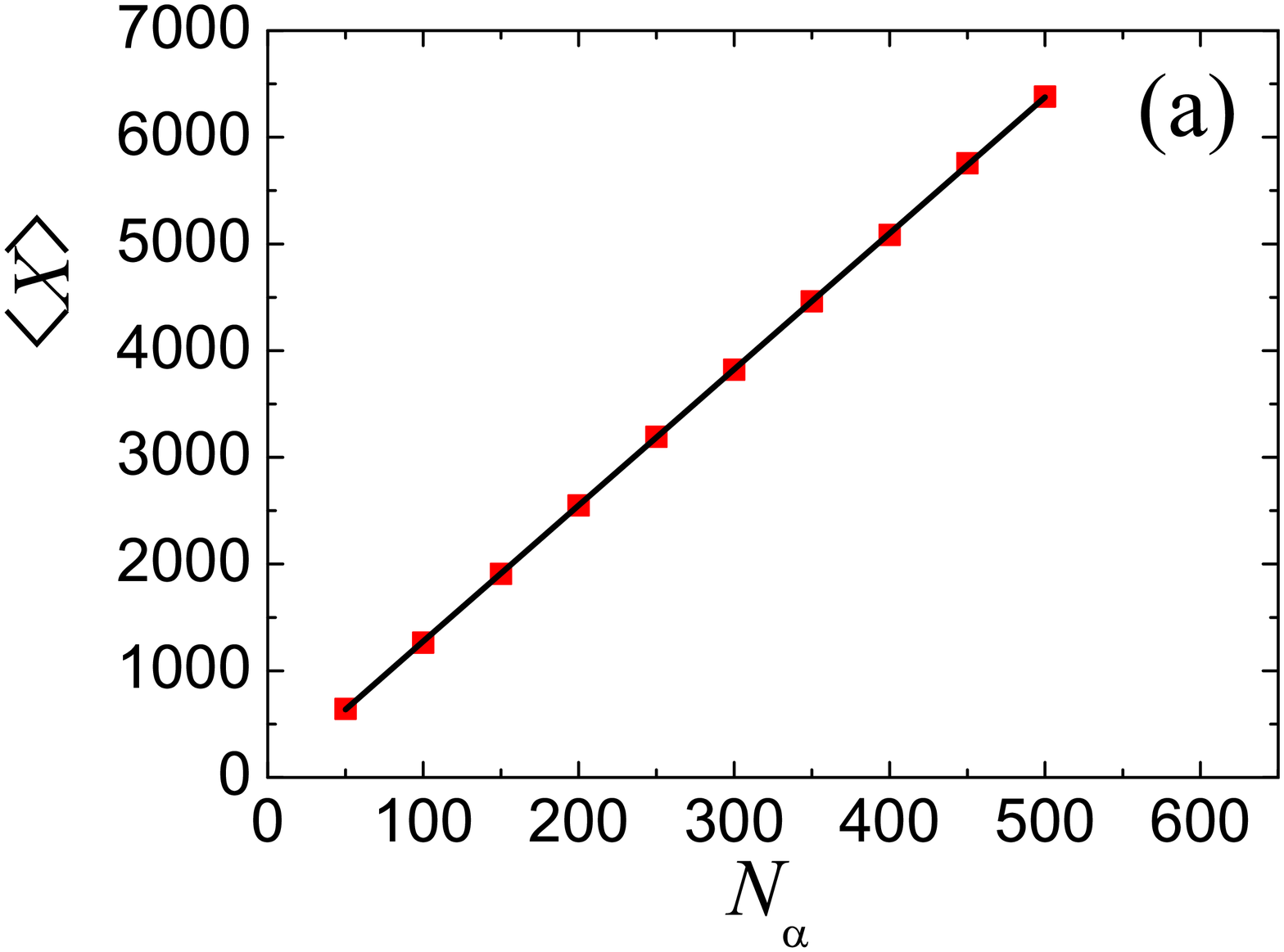}}\quad
    \subfigure{\includegraphics[width=3in]{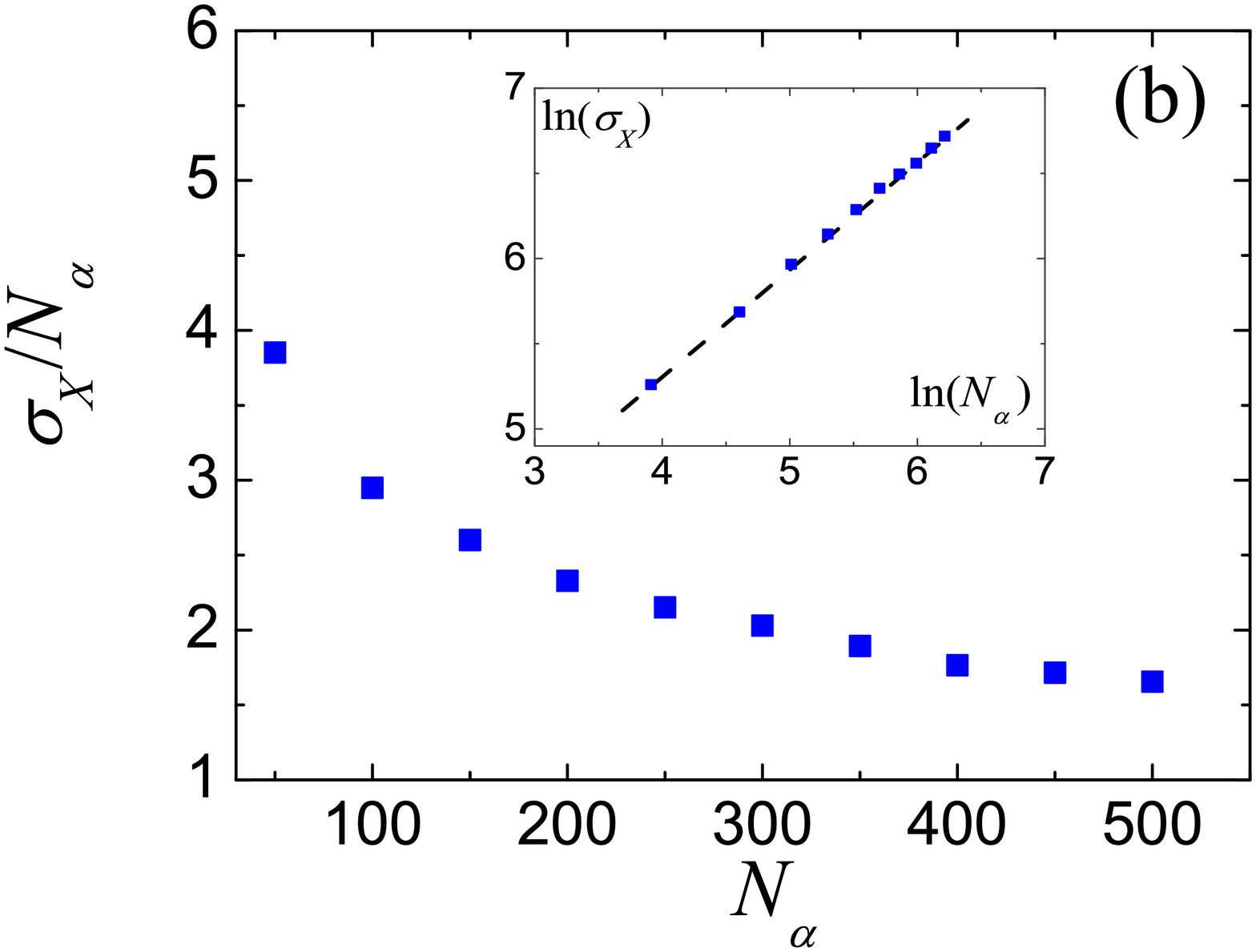}}
     }
\caption{The means (red squares) and standard deviations (blue squares) associated with cross-links in a 
two-network model with $\protect\kappa _{\protect\alpha }=25$ and 
$\protect\chi _{\protect\alpha }=0.5$, as a function of 
$N_\alpha=N_{1}=N_{2}$. In (a), $N_\alpha\kappa_\alpha/2$ is also plotted (solid line) for comparison. The inset in (b) shows a log-log plot for $\sigma_\alpha$ (blue squares) as a function of $N_\alpha$. For comparison, the dashed line reflects a power law, $\propto N_\alpha^{0.63}$.}
\label{diffN}
\end{figure*}

Finally, we vary $L=N_{1}=N_{2}$ while holding $\kappa _{\alpha }=\kappa =25$
and $\chi _{\alpha }=\chi =0.5$. Due to limited computing power, we only
explored an order of magnitude around $100$: $L\in \left[ 50,500\right] $,
the results of which are shown in Fig.~\ref{diffN}. Again, the dependence of
the mean, $\langle X\rangle $, is easy to understand, namely, $\kappa \chi
L\simeq 12.5L$ (solid line). However, similar to the systems reported above,
the behavior of $\sigma _{X}$ is more intriguing. In an inset of Fig.~\ref
{diffN}(b), we show a log-log plot of $\sigma _{X}$-$L$ and see that $\sigma
_{X}$ scales well as $L^{0.63}$. Such anomalous scaling hints at critical
phenomena and deserves a thorough investigation. In particular, the
incidence matrix associated with the two communities can be viewed as an $%
L\times L$ Ising model (in the lattice gas language, with $0,1$ in the
entries). Then, $X$ maps into $2M-L^{2}$, with $M$ being the total
magnetization, while the variance $\sigma _{X}^{2}$ corresponds to the Ising
susceptibility ($\times L^{2}$). So, our findings here imply a \textit{%
decreasing} `susceptibility' ($\sim L^{-0.74}$). This remarkable
property can be argued qualitatively, as follow. Note that $\kappa \chi $
controls the creation of a cross-link, so that any particular link occurs,
roughly, with probability $\kappa \chi /L$. Thus, increasing $L$ while
holding $\kappa \chi $ fixed corresponds to an increasingly stronger
magnetic field, which in turn, leads to a decreasing magnetic
susceptibility. In the next paper of this series \cite
{LiuBasslerSchmittmannZia14}, such a mapping can be established analytically
(for the two-population model with maximally opposite preferences), while a
system with $N_{1}=N_{2}$ can be interpreted as a peculiar critical point 
\cite{LiuSchmittmannZia12}.

\subsubsection{Results from varying $N_{\protect\alpha }$ and $\protect%
\kappa _{\protect\alpha }$.}

\begin{figure*}[tbp]
\centering
\mbox{
    \subfigure{\includegraphics[width=3in]{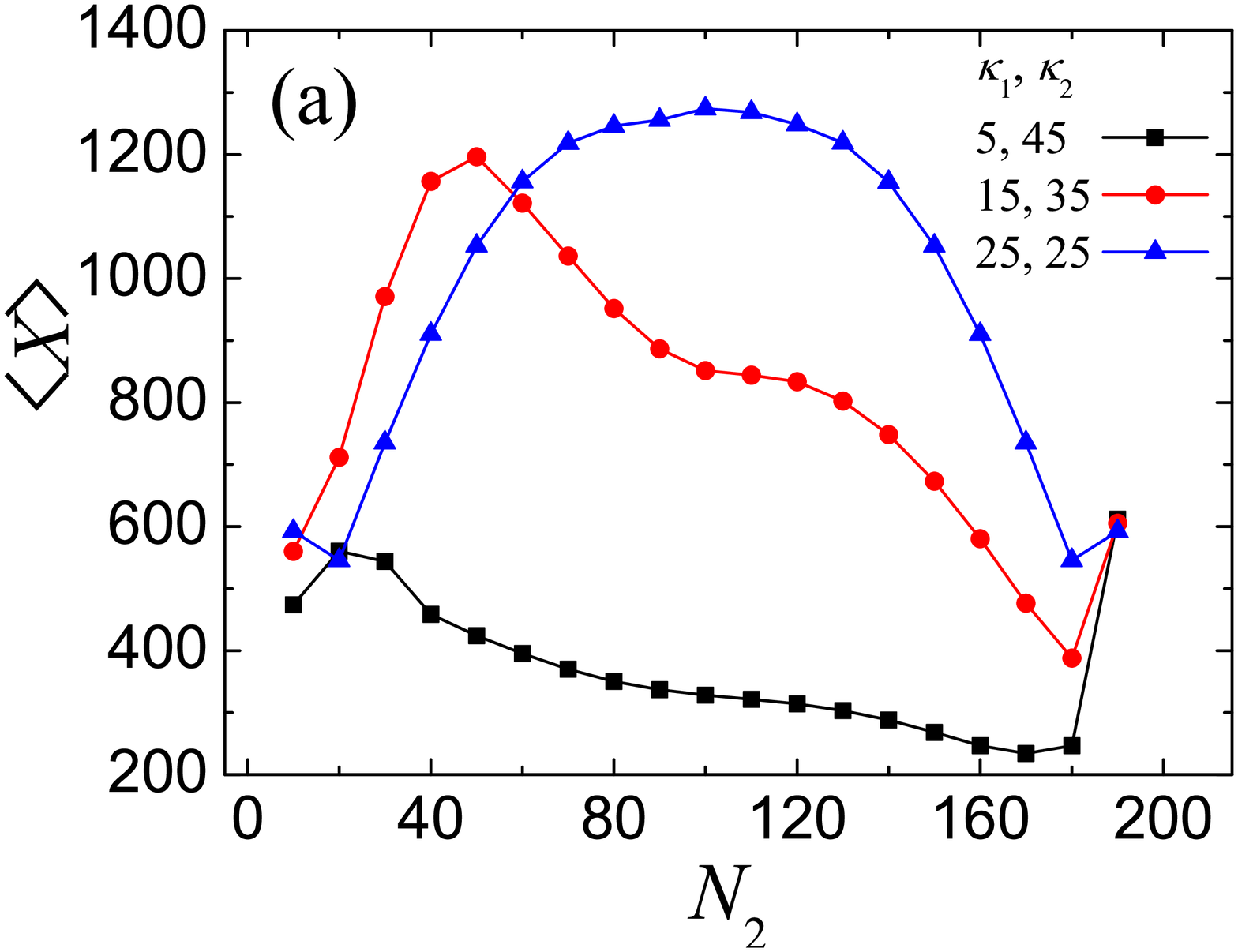}}\quad
    \subfigure{\includegraphics[width=3in]{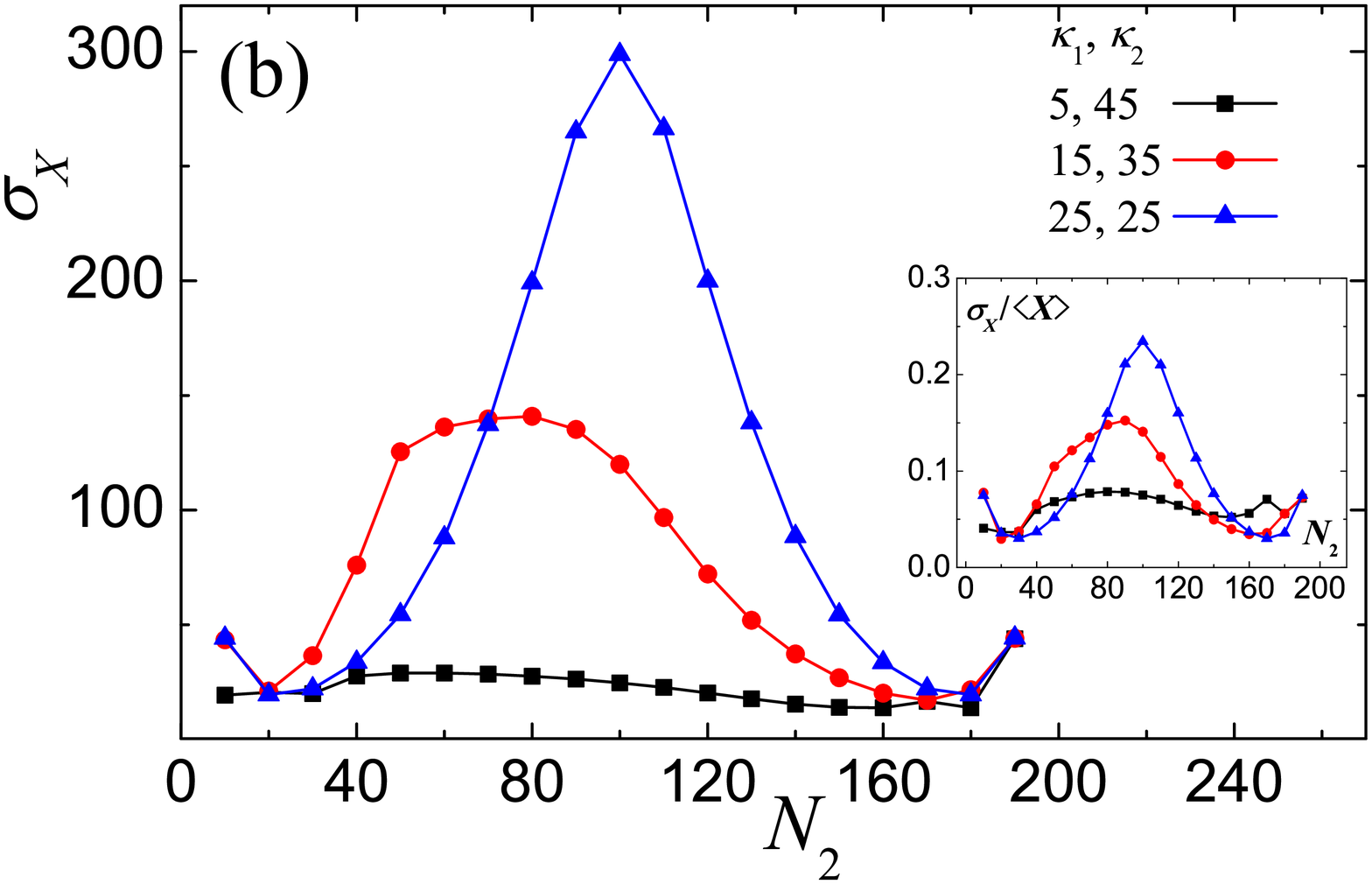}}
     }
\caption{The means (a) and standard deviations (b) associated with 
cross-links in a two-network model with 
$\protect\chi _{\protect\alpha }=0.5$,
$N_1+N_2=200$, $\protect\kappa _1+\protect\kappa _2=50$, 
as a function of $N_{2}$, for various $\protect\kappa$. The inset in (b) shows a plot of $\sigma_X/\langle X\rangle$ versus $N_2$.}
\label{general}
\end{figure*}

In the previous subsection, we explored the region around a special
symmetric two-network system by varying one pair of parameters (i.e., along
certain lines). Here we turn our attention to a more general case, in which
the networks differ by two parameters, specifically, just the sizes and
the preferences. To make comparisons with the previously studied systems, we
restrict ourselves to systems with fixed sums: $N_{1}+N_{2}=200$, $\kappa
_{1}+\kappa _{2}=50$, and $\chi _{1}+\chi _{2}=1$. Moreover, from the
results above (especially Fig.~\ref{diffS}), we see that the effects of
changing $\chi $ are minimal. Thus, we simply fix both $\chi $'s to $0.5$.
As we vary $N_{\alpha }$, over the entire range, we need to keep, say, only $%
\kappa _{1}\leq \kappa _{2}$ to access the whole subspace of interest. This
choice allows us to refer to $\alpha =1,2$ as the introverts and extroverts,
respectively. In Fig.~\ref{general}, we provide results for $\langle
X\rangle $ and $\sigma _{X}$ as a function of $N_{2}$, for three different
pairs of $\left( \kappa _{1},\kappa _{2}\right) $. We choose ($5,45$), ($%
15,35 $), and ($25,25$) partly for convenience and partly for having $\kappa
_{1}$'s ratios at $1:3:5$.

Focusing on the last pair first (blue points), we note that both curves are
symmetric around $N_{2}=100$. Since the simulations were performed for pairs
of $N_{2}$ around 100, the observed symmetry is an indication of the level
of our statistical errors. The values of $\langle X\rangle $ and $\sigma
_{X} $ at the center are entirely consistent with the findings shown above:
approximately $1300$ and $300$ respectively. Away from the center, $\langle
X\rangle $ appears to decrease slowly at first, but turns up when $N_{2}$
reaches to $\sim 10\%$ of the boundaries. The slow decrease is related
to $N_{1}N_{2}=\left( 200-N_{2}\right) N_{2}$, the maximum allowed value for 
$X$. Indeed, in this regime, the fraction 
\begin{equation}
f\equiv \langle X\rangle /N_{1}N_{2}
\end{equation}
hovers around 13\%, i.e., $\sim \kappa \chi $ divided by the average
number of nodes in each community. On the other hand, the non-monotonic
behavior can hardly be expected. In the next subsection, we will present an
approximation scheme which can provide some insight into this remarkable
phenomenon.

Turning to the two asymmetric cases, we are not surprised by the lack of
symmetry in the curves. However, there are prominent and interesting
features, the origins of which do not readily come to mind. First, the
levels of $\langle X\rangle $ are generally reduced, despite both $\chi $'s
being held at 0.5 and the average preference remaining at 25. Specifically,
for a wide range of $N_{2}\gtrsim 100$, the $f$'s are relatively constant,
matching roughly the ratio $1:3:5$. Thus, it appears that the introverts are
controlling the level of cross-links. Why the extroverts play a lesser role
may be argued as follows: When the number of extroverts is well above $
\kappa _{2}$, they can be `content' by maintaining more links to other
extroverts, instead of adding cross-links. Of course, it would be highly
desirable to formulate an analytic and more convincing approximation scheme.
Second, in addition to the sharp upturn for $N_{2}\sim 200$, the peaks of $%
\langle X\rangle $ are shifted to smaller $N_{2}$, to approximately $20$
and $50$ for $\kappa _{1}=5$ and $15$ respectively. These features can also
be roughly argued. Given that the preferred degrees are generally less than
the number of nodes, it is understandable that the introverts are more
likely to be frustrated, by the eagerness of extroverts to make cross-links.
As we decrease $N_{2}$, (especially to values below $N_{1}$) we can expect
such frustration to decrease, leading to the rise in $\langle X\rangle $
observed. Arguably, an `optimal' state might be characterized by a balance
between the total number of cross-links the introverts prefer, $N_{1}\kappa
_{1}$, and that of the extroverts, $N_{2}\kappa _{2}$. (Note that we can
ignore $\chi $ for this rough argument, since it affects predominantly the \textit{%
rate} of actions on cross-links). This balance occurs at $N_{2}=4\kappa _{1}$
for our parameters, i.e., $20$ and $60$ for the runs with black and red
points, respectively. While such arguments produce a rough understanding of
the data, more quantitative improvements are clearly needed.

Lastly, we turn to the fluctuations in the cross-links, characterized here
by only the standard deviation $\sigma _{X}$, see Fig.~\ref{general}(b). As
pointed out above, these are far larger than naively expected and
explaining their presence remains a challenge. Here, we merely highlight what
is in the figure: the asymmetric curves for $\kappa _{1}\neq \kappa _{2}$,
the variation by over an order of magnitude, and the `calming influence of
the more introverted.' The last of these comments refers to the almost
constant $\sigma _{X}$ for the $\kappa _{1}=5$ case. We end this subsection
with another observation. Noting that the peaks of $\sigma _{X}$ for the
asymmetric $\kappa $'s are also displaced toward the peaks of $\left\langle
X\right\rangle $, we plot a `normalized standard deviation,' $\sigma
_{X}/\left\langle X\right\rangle $, see inset in Fig.~\ref{general}(b). It is interesting that the peaks here
are now located much closer to the center (though the curves remain
asymmetric). Perhaps a detailed analysis of this quantity will facilitate
the formulation of a viable theory.

\subsubsection{Theoretical understanding of $X$ in the $N_{1}\ll N_{2}$
regime.}

\begin{figure}[tbp]
\centering
\includegraphics[width=3.5in]{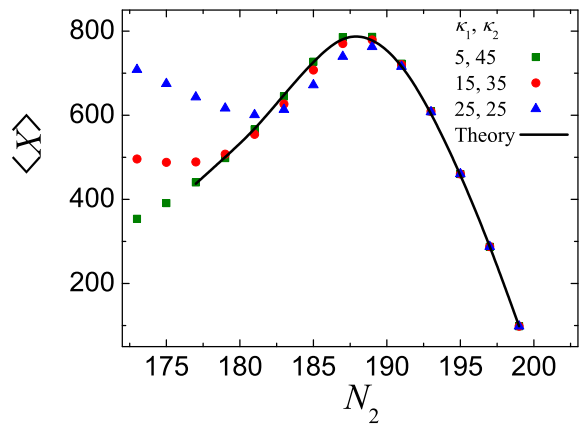}
\caption{This figure shows $\langle X \rangle$ of systems in the $N_{1}\ll N_{2}$ regime. The associated system parameters are $\protect\chi _{\protect\alpha }=0.5$,
$N_1+N_2=200$, $\protect\kappa _1+\protect\kappa _2=50$. 
For this parameter set, the regime with $N_2$ close to $200$ corresponds to the $N_{1}\ll N_{2}$ regime.}
\label{distinctX}
\end{figure}

In this last subsection, we present an approximation scheme which provides
reasonably good agreement with data in a special regime. It is clear that,
in a typical point in parameter space, there are so many competing features
in our model that the contributions of all these factors will be difficult to
untangle. A remarkable phenomenon -- the upturn of $\left\langle
X\right\rangle $ as $N_2$ nears its upper bound in Fig.~\ref{general}(a) --
is observed. Moreover, it appears that $%
\left\langle X\right\rangle $ assumes the same value at $N_2=190$ for all
three pairs of $\kappa $'s! It behooves us, therefore, to explore this
regime in more detail. The result, shown in Fig.~\ref{distinctX}, hints at
the existence of some underlying `universal' properties. Focusing on this
regime, we discover that a simplification emerges, allowing us to gain some insight into this universality.

In the regime of interest, introverts are seriously outnumbered and highly
frustrated. In other words, they find themselves with far more contacts
than they prefer (i.e., $k\gg \kappa _1$), so that, when selected to act, they
will always cut a link. A more precise and general characterization of this
regime is $N_1\kappa _1\chi _1\ll N_2\kappa _2\chi _2$. The result of this
activity is that there are no (or extremely few) $I$-$I$ links in the
system, so that we can attempt an approximation scheme for $\rho _1^{ss}\left(
k\right) $, the degree distribution of an $I$.\footnote{%
Note that, since we have assumed $k_{11}=0$, we can drop the subscripts in $%
k_{12}$, while $\rho _{12}\left( k\right) $ is also identical to $\rho
_{1}\left( k\right) $.} From $\rho _1^{ss}$, the average $\left\langle
k\right\rangle $ can be computed. Since every link is a cross-link, $%
\left\langle X\right\rangle $ is just $N_1\left\langle k\right\rangle $.

\begin{figure*}[tbp]
\centering
\mbox{
    \subfigure{\includegraphics[width=3in]{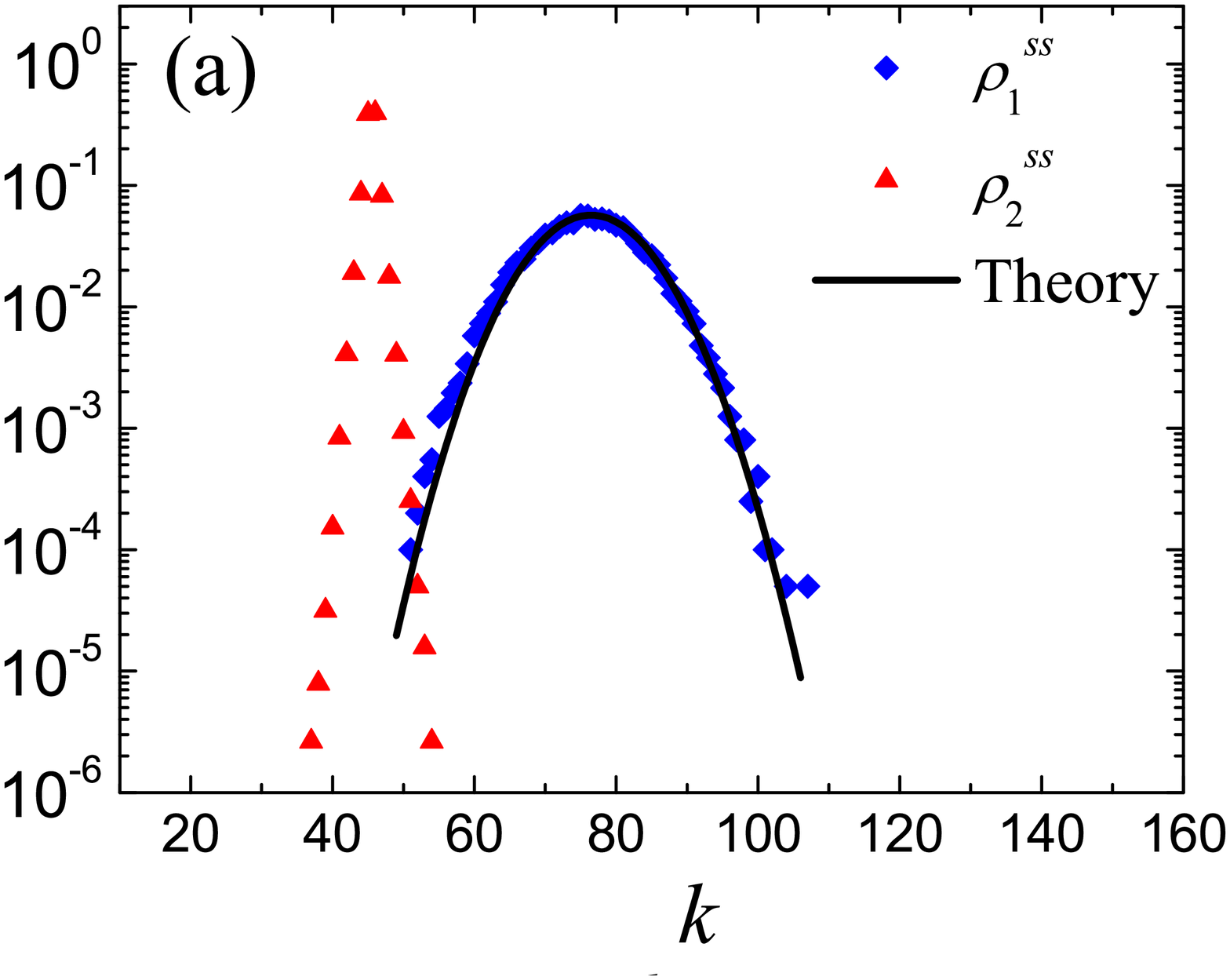}}\quad
    \subfigure{\includegraphics[width=3in]{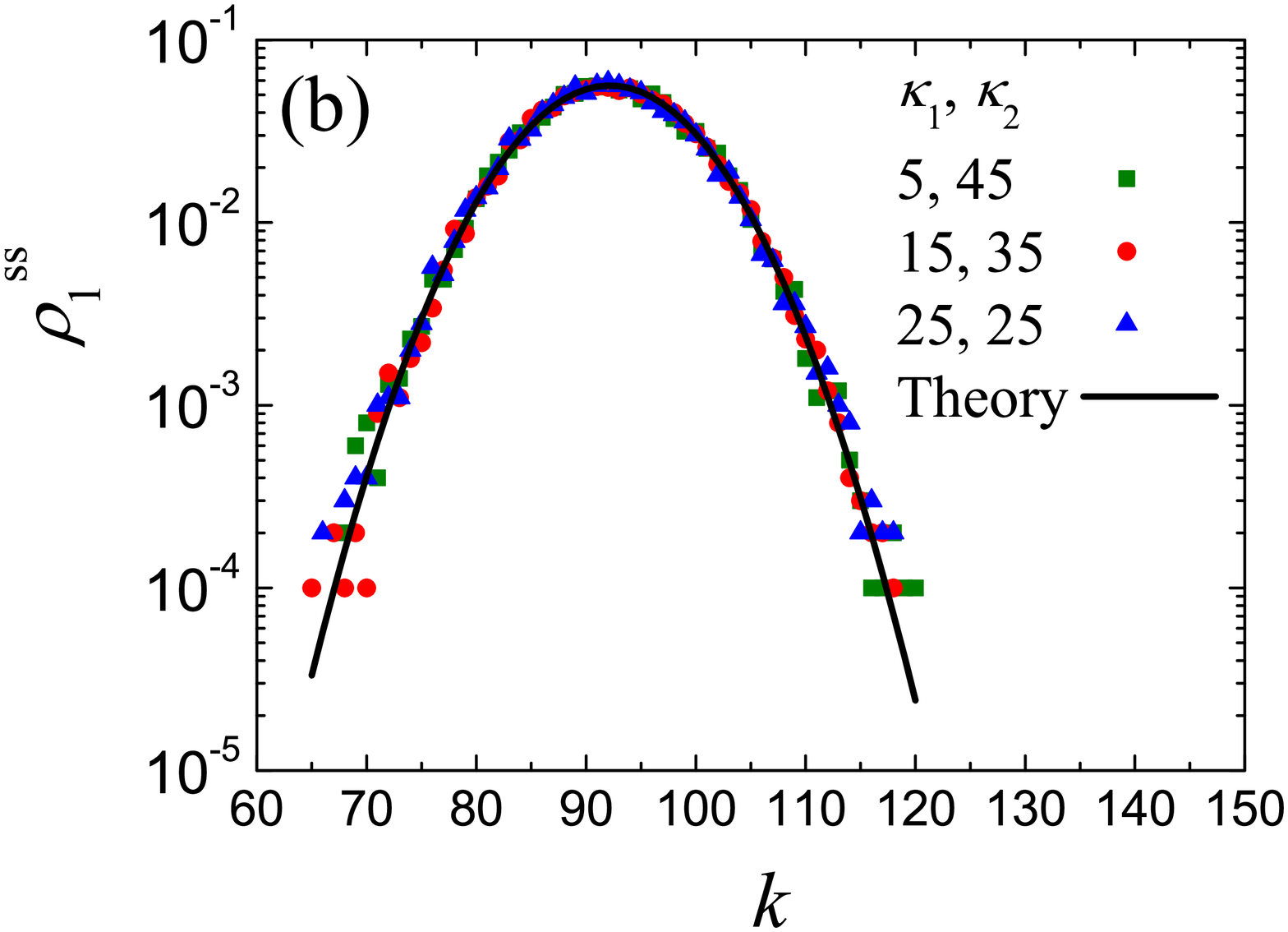}}
     }
\caption{(a) The total degree distributions for a two-network model 
($N_1=10$, $N_2=190$, $\protect\kappa _1=5$, $\protect\kappa _2=45$ 
and $\protect\chi _\protect\alpha =0.5$): $\protect\rho _1^{ss}$ 
(blue diamonds) and $\protect\rho _2^{ss}$ (red triangles). (b) The total degree distributions $\protect\rho _1^{ss}$ for a two-network model with three different values of $\kappa_1$, $\kappa_2$. The other associated parameters are $N_1=5$, $N_2=195$, and $\protect\chi _\protect\alpha=0.5$. In both figures, solid lines represent theoretical predictions.}
\label{typicalDD}
\end{figure*}

The strategy here is the same as the one used for the single network:
finding the steady state $\rho ^{ss}$ by balancing the rates for increasing
and decreasing $k$, namely, 
\begin{equation}
\rho ^{ss}(k)W[k\rightarrow k+1]=\rho ^{ss}(k+1)W[k+1\rightarrow k]
\label{rhoW}
\end{equation}
where $W[k\rightarrow k^{\prime }]$ specifies the probability for a node
with degree $k$ to become $k^{\prime }$. To find the appropriate $W$'s here,
we rely on the following argument. Focus on a particular introvert, $i$,
with existing degree $k$. For the regime of interest, we can assume $%
k>\kappa _1$, so that the only way for it to gain a link is for an extrovert
(not already connected to $i$) choosing to add a cross-link to it. Several
factors contribute to the rate for such a process:

\begin{itemize}
\item $\left( N_2-k\right) /N$, the probability for an extrovert \textit{not}
connected to $i$ to be selected

\item $w_2^{+}$, the probability that this extrovert will add a link

\item $\chi _2$, the probability for it to add a cross-link ($1/2$ here), and

\item $p$, the probability that this cross-link is added to $i$.
\end{itemize}

Now, in this regime, we believe the extroverts should be mostly `content' ($%
\kappa _{2}\ll N$). This belief is supported by the observed degree
distributions, a typical case shown as red points in Fig.~\ref{typicalDD}(a).
Thus, we will approximate $w_{2}^{+}$ by $1/2$. As for the last quantity, $%
p^{-1}$ should be the number of introverts unconnected to our extrovert. We
may estimate it by $N_{1}\left( 1-f\right) $, where $f$ is the fraction of
cross-links that are present. To be slightly more accurate, especially crucial for
small $N_{1}$, we recognize that our extrovert is already connected to $i$
and so, we propose 
\begin{equation}
\frac{1}{p}\simeq 1+\left( N_{1}-1\right) \left( 1-f\right) .
\end{equation}%
Putting the factors together, we have 
\begin{equation}
W[k\rightarrow k+1]=\frac{N_{2}-k}{N}w_{2}^{+}\chi _{2}p\simeq \frac{\left(
N_{2}-k\right) /4N}{1+\left( N_{1}-1\right) \left( 1-f\right) }.
\end{equation}%
On the other hand, the contributions to $W[k+1\rightarrow k]$ come from two
processes. One is node $i$ being selected to cut a cross-link. The probability of selecting this particular node $i$ is $1/N$, and with rate $1$ it will cut a link. But this action will be taken on a cross-link only with probability $%
\chi _{1}$ ($1/2$ here). In the other process, an extrovert connected to $i$
is chosen (probability $k/N$) and cuts one of its cross-links (probability $%
w_{2}^{-}\chi _{2}\simeq 1/4$). Following the argument above for $p^{-1}$, we
have 
\begin{equation}
\frac{1}{p}\simeq 1+\left( N_{1}-1\right) f.
\end{equation}%
Of course, here $p$ corresponds to the probability that the extrovert \textit{cuts} the cross-link to $i$. Therefore, we reach at 
\begin{equation}
W[k+1\rightarrow k]=\frac{1}{2N}+\frac{k+1}{N}w_{2}^{-}\chi _{2}p\simeq 
\frac{1}{2N}+\frac{\left( k+1\right) /4N}{1+\left( N_{1}-1\right) f}.
\end{equation}%
With explicit expressions for the $W$'s, we exploit Eqn.~(\ref{rhoW}) to
derive a recursion relation for the steady state degree distribution: 
\begin{equation}
\rho _{1}^{ss}\left( k+1\right) =\rho _{1}^{ss}\left( k\right) R\left(
k\right)   \label{DD-rr}
\end{equation}%
where 
\begin{equation}
R\left( k\right) =\left\{ \frac{N_{2}-k}{1+\left( N_{1}-1\right) \left(
1-f\right) }\right\} \left\{ 2+\frac{k+1}{1+\left( N_{1}-1\right) f}\right\}
^{-1}  \label{R}
\end{equation}%
Thus, $\rho _{1}^{ss}\left( k\right) $ is explicitly 
\begin{equation}
\rho _{1}^{ss}\left( k\right) =\rho _{1}^{ss}\left( 0\right) \prod_{\ell
=0}^{k-1}R\left( \ell \right)   \label{rho-ans}
\end{equation}%
where $\rho _{1}^{ss}\left( 0\right) $ can be fixed by the normalization $\Sigma
_{0}^{N_{2}}\rho _{1}^{ss}\left( k\right) =1$.

So far, $f$ is not a known quantity. But it can determined self
consistently, through the equations 
\begin{equation}
f=\frac{\left\langle X\right\rangle }{N_{1}N_{2}}=\frac{\left\langle
k\right\rangle }{N_{2}}=\frac{1}{N_{2}}\sum_{0}^{N_{2}}k\rho _{1}^{ss}\left(
k\right)   \label{f}
\end{equation}%
The result is a \textit{prediction} (i.e., no fitting parameters)
for $\rho _{1}^{ss}$ and so, $\left\langle X\right\rangle $. Plotted as
black lines in Figs.~\ref{typicalDD} and \ref{distinctX}, these predictions
agree remarkably well -- provided we remain in the $N_{1}\ll N_{2}$ regime.
Clearly, the deviations of $\left\langle X\right\rangle $ from the
theoretical curve in Fig.~\ref{distinctX} reflect the limits of this regime.

Finally, we return the issue of `universality.' Note that results Eqns.~(\ref
{R},\ref{rho-ans},\ref{f}) are \textit{independent} of the $\kappa $'s. Of
course, their validity relies on the assumption that $\left\langle
k_{1}\right\rangle \gg \kappa _{1}$, which seems reasonable in these cases
where all the introverts are highly frustrated. In Figs.~\ref{typicalDD}(b),
we show data for a more extreme system ($N_{1},N_{2}=5,195$) in which this
notion of universality is indeed well borne out. An alternative perspective
is that, in this regime, the $N_{1}\times N_{1}$ block of the full $N\times N
$\ adjacency matrix is frozen at zero, so that much of what we wish to
compute can be gleaned from the smaller $N_{1}\times N_{2}$  incidence
matrix. As long as we restrict ourselves to considering `zero $N_{1}\times
N_{1}$ blocks,' the role of $\kappa _{1}$ is entirely marginal.

\section{SUMMARY AND OUTLOOK}

In this paper, we present further explorations of preferred degree networks
and their interactions in a more systematic Monte Carlo study. Specifically,
we consider the effects of just one way of coupling the two networks,
through $\chi $, the probability that a node in one network adds or cuts a
link to a partner in the other community. Thus, $\chi =0$ corresponds to
two completely decoupled networks, while a system with $\chi =1$ can be
regarded as `maximally coupled.' Even restricting ourselves to this simple
interaction, we are faced with a large, 6-dimensional parameter space:
the number of nodes in each network ($N_{1,2}$), their preferred degrees ($\kappa
_{1,2}$), and their couplings $\chi _{1,2}$. We began with a study of the most
symmetric system Eqn.~(\ref{Symmetric}), consisting of two \textit{identical}
networks coupled by $\chi =1/2$. With limited computational resources, we
ventured from this point onto several 1-d and 2-d subspaces. Here, the
preferred degrees are typically different and we chose the convention $%
\kappa _{1}\leq \kappa _{2}$, referring to them as `introverts' and
`extroverts.' Simulating these systems and letting them settle into steady
states, we characterize them by measuring various degree distributions as
well as the behavior of $X$, the total number of cross-links between the
communities. While some results are expected, other properties
are quite surprising. 

One remarkable result is the drastically different behavior between two very
similar systems, one being the completely symmetric two-network system Eqn.~(\ref{Symmetric}) and the other being a single homogeneous network (with the
same total population) partitioned into identical halves. In particular,
defining $X$ for the homogeneous network as the links between these two
halves, we find that its stationary distribution, $P^{ss}(X)$, is well
described by a Gaussian distribution, with an easily predicted mean, $%
\left\langle X\right\rangle $, and standard deviation, $\sigma _{X}$.
However, in simulations of the symmetric two-network system, $P^{ss}(X)$
displays a very broad and flat plateau. The standard deviation here is an
order of magnitude larger and so far, understanding it remains a challenge.
The different behaviors indicate that, despite its simplicity, this way of
coupling two networks has a profound effect on the system.

Away from the symmetric systems, most features of the simulation results for 
$\left\langle X\right\rangle $ and $\sigma _{X}$ can be qualitatively
understood, although a good theory will be needed to provide acceptable
quantitative agreements. Remarkably, within the regions we explored, we
observed `universal' behavior in asymmetric systems when the introverts are
far outnumbered, in the following sense. Not only $\left\langle
X\right\rangle $ and $\sigma _{X}$, but also the full degree distribution
for the introverts, become \textit{independent} of the $\kappa $'s. Insight
to this behavior can be found by noting that, in this regime, the introverts
are so frustrated that their only action is cutting links. As a result,
there are no links between the introverts and the state of each can be
specified by the number of cross-links alone. An approximation scheme for
their degree distribution can be formulated, leading to very successful
predictions. 

These findings, though in a rather limited region of control parameter
space, reveal many non-trivial phenomena in a system with just two networks,
coupled in a simple way. Quantitative explanations of much of the data are
still lacking. To make progress, we may extend the same approximation
schemes to study the joint distributions $P_{\alpha }\left( k_{\alpha \alpha
},k_{\alpha \beta }\right) $. Preliminary analysis indicates that, unlike the
exact transition rates for the full, microscopic distribution, such a set
of approximate rates obeys `local' detailed balance, i.e., the only
irreversible Kolmogorov loops are those around the $k=\kappa
_{\alpha }$ line. Thus, it may be possible for $P_{\alpha }^{ss}$ to be
found analytically. An in-depth study is underway. Meanwhile, along the
lines of our successful theory for the special regime here (where the
introverts are highly frustrated), we can consider the case in which both
parties are `maximally frustrated.' With e\underline{x}treme \underline{i}%
ntroverts ($\kappa _{1}=0$) and \underline{e}xtroverts ($\kappa _{2}=\infty $%
) in our system, we coined it the `XIE model.' Within a short time, the
intra-community links will be frozen (empty and full, respectively) while
only the cross-links are dynamic. The $N\times N$ adjacency matrix reduces
fully to the $N_{1}\times N_{2}$ incidence matrix and our problem simplifies
considerably. Even in this extreme case, surprising behavior emerges, some
of which has been reported \cite{LiuSchmittmannZia12}. Thanks to the
restoration of detailed balance, we can solve the master equation and find
the microscopic stationary distribution exactly \cite{LiuSchmittmannZia12}. The next paper of this
series \cite{LiuBasslerSchmittmannZia14} will be devoted to more systematic simulations as well as more
in-depth analytic studies.

\ack

We thank K. Bassler, S. Jolad, L.B. Shaw and Z. Toroczkai for illuminating
discussions. This research is supported in part by the US National Science
Foundation, through grant DMR-1244666.


\section*{References}

\begin{thebibliography}{99}

\bibitem{LiuJoladSchZia13} Liu W, Jolad S, Schmittmann B and Zia R K P 2013 
\textit{J. Stat. Mech. Theory Exp.} \textbf{2013} P08001

\bibitem{Strogatz01} Strogatz S H 2001 \textit{Nature} \textbf{410} 268

\bibitem{AlbertBarabasi02} Albert R and Barab\'{a}si A-L 2002 \textit{Rev.
Mod. Phys.} \textbf{74} 47

\bibitem{DorogovtsevMendes02} Dorogovtsev S N and Mendes J F F 2002 \textit{%
Adv. Phys.} \textbf{51} 1079

\bibitem{Newman03} Newman M E J 2003 \textit{SIAM Rev.} \textbf{45} 167

\bibitem{EstradaFoxHighamOppo10} Estrada E, Fox M, Higham D and Oppo G-L
(eds) 2010 \textit{Network Science:\ Complexity in Nature and Technology}
(Springer, New York)

\bibitem{WattStrogatz98} Watts D J and Strogatz S H 1998 \textit{Nature
(London)} \textbf{393} 440

\bibitem{AlbertJeongBarabasi99} Albert R, Jeong H and Barab\'{a}si A-L 1999 
\textit{Nature (London)} \textbf{401} 130

\bibitem{BarabasiAlbert99} Barab\'{a}si A-L and Albert R 1999 \textit{Science%
} \textbf{286} 509

\bibitem{BarratBarthelemyVespignani08} Barrat A, Barthelemy M and Vespignani
A 2008 \textit{Dynamical processes on complex networks} (Cambridge
University Press)

\bibitem{DorogovtsevGAMJ08} Dorogovtsev S N, Goltsev A V and Mendes J F F
2008 \textit{Rev. Mod. Phys.} \textbf{80} 1275

\bibitem{GrossDCBB06} Gross T, D'Lima C J D and Blasius B 2006 \textit{Phys.
Rev. Lett.} \textbf{96} 208701

\bibitem{GrossBlasius08} Gross T and Blasius B 2008 \textit{J. R. Soc.
Interface} \textbf{5} 259

\bibitem{RinaldiPeerenboomKelly01} Rinaldi S, Peerenboom J and Kelly T 2001 
\textit{IEEE Contr. Syst. Mag.} \textbf{21} 11

\bibitem{PanzieriSetola08} Panzieri S and Setola R 2008 \textit{Int. J.
Model. Ident. Contr.} \textbf{3} 69

\bibitem{Vespignani10} Vespignani A 2010 \textit{Nature (London)} \textbf{464%
} 984

\bibitem{BuldyrevParshaniPaulStanleyHavlin10} Buldyrev S V, Parshani R, Paul
G, Stanley H E and Havlin S 2010 \textit{Nature (London)} \textbf{464} 1025

\bibitem{BuldyrevShereCwilich11} Buldyrev S V, Shere N W and Cwilich G A
2011 \textit{Phys. Rev. E} \textbf{83} 016112

\bibitem{KurantThiran06} Kurant M and Thiran P 2006 \textit{Phys. Rev. Lett.}
\textbf{96} 138701

\bibitem{E-R} Erd\H{o}s P and R\'{e}nyi A 1959 \textit{Pub. Math.} \textbf{6}
290

\bibitem{PlatiniZia10} Platini T and Zia R K P 2010 \textit{J. Stat. Mech.
Theory Exp.} \textbf{2010} P10018

\bibitem{ZiaLiuJoladSchmitt11} Zia R K P, Liu W, Jolad S and Schmittmann B
2011 \textit{Physics Procedia} \textbf{15} 102

\bibitem{JoladLiuSchmittmannZia12} Jolad S, Liu W, Schmittmann B and Zia R K
P 2012 \textit{PLoS ONE} \textbf{7(11)} e48686

\bibitem{LiuSchmittmannZia12} Liu W, Schmittmann B and Zia R K P 2012 
\textit{EPL} \textbf{100} 66007

\bibitem{ZS2007} Zia R K P and Schmittmann B 2007 \textit{J. Stat. Mech.
Theory Exp.} \textbf{2007} P07012

\bibitem{LiuBasslerSchmittmannZia14} Liu W, Bassler K E, Schmittmann B, and
Zia R K P to be published.

%\bibitem{ZanetteR.JBioPhys.2008}
%Zanette D H and Risau-Gusm{\'a}n S 2008 {\it Journal of biological physics} {\bf 34} 135

%\bibitem{GrossK.EPL.2008}
%Gross T, and Kevrekidis I G 2008 {\it EPL} {\bf 82} 38004

%\bibitem{ShawS.PRE.2008}
%Shaw L B and Schwartz I B 2008 {\it Phys. Rev. E} {\bf 77} 066101

%\bibitem{SchwarzkopfRM.PRE.2010}
%Schwarzkopf Y, R\'akos A and Mukamel D 2010 {\it Phys. Rev. E} {\bf 82} 036112

%\bibitem{MarceauPHAD.PRE.2010}
%Marceau V, No\"el P A, H\'ebert-Dufresne L, Allard A and Dub\'e L J 2010 {\it Phys. Rev. E} {\bf 82} 036116

%\bibitem{WangCSA.JPhysA.2011}
%Wang B, Cao L, Suzuki H and Aihara K 2011 {\it Journal of Physics A: Mathematical and Theoretical} {\bf 44} 035101

%\bibitem{http://www.moe.gov.sg/media/press/2003/pr20030326.htm}
%http://www.moe.gov.sg/media/press/2003/pr20030326.htm

%\bibitem{Kolmogorov} 
%Kolmogorov A N 1936 {\it Math Ann.} {\bf 112} 155

%\bibitem{ParshaniBuldyrevHavlin10}
% R. Parshani, S. V. Buldyrev, and S. Havlin, Phys. Rev. Lett. {\bf 105}, 048701 (2010).

%\bibitem{Newman01}
% M. E. J. Newman, Proc. Natl. Acad. Sci. U S A. {\bf 98}, 404 (2001).

%\bibitem{Eubank04}
% S. Eubank, H. Guclu, V. S. A. Kumar, M. V. Marathe, A. Srinivasan, Z. Toroczkai, and N. Wang, Nature (London) {\bf %429}, 180 (2004).

%\bibitem{GonzalezHidalgoBarabasi08}
%M. C. Gonzalez, C. A. Hidalgo and A-L. Barab\'{a}si, Nature (London) {\bf 453}, 779 (2008).

%\bibitem{Volz04}
%E. Volz, Phys. Rev. E. {\bf 70}, 056115 (2004).

%\bibitem{BenczikSchmitZia08}
% I. J. Benczik, S. Z. Benczik, B. Schmittmann, and R. K. P. Zia, EPL {\bf 82}, 48006 (2008).

%\bibitem{BenczikSchmitZia09}
%I. J. Benczik, S. Z. Benczik, B. Schmittmann, and R. K. P. Zia, Phys. Rev. E {\bf 79}, 046104 (2009).
%\bibitem{MBTI} 
%http://www.capt.org/mbti-assessment/estimated-frequencies.htm.
\end{thebibliography}
\end{document}